\documentclass[11pt,oneside]{amsart}
\usepackage{latexsym,amssymb,amsmath,youngtab}
\usepackage{amsmath,amsthm,amssymb,amscd}%,MnSymbol} ciken: MnSymbol distorts many beautiful symbols, including but not limited to \subset and \simeq.
\usepackage[mathscr]{eucal}
\usepackage{verbatim}
\usepackage{ulem}
\usepackage{url}
\usepackage{color}
\usepackage{algorithmic} %, algorithmic-fix}
\usepackage{listings, algorithm}

\newcommand{\codim}{{\rm codim\ }}

\newtheoremstyle{custom}% name
  {3pt}%      Space above
  {3pt}%      Space below
  {\slshape}%         Body font
  {}%         Indent amount (empty = no indent, \parindent = para indent)
  {\bfseries}% Thm head font
  {.}%        Punctuation after thm head
  { }%     Space after thm head: " " = normal interword space;
   {}%         Thm head spec (can be left empty, meaning `normal')
\theoremstyle{custom}
\newtheorem{theorem}{Theorem}[subsection]

\newtheorem{proposition/definition}[theorem]{Proposition/Definition}

\theoremstyle{definition}

\newtheorem{example}[theorem]{Example}

\theoremstyle{remark}
\newtheorem{remark}[theorem]{Remark}

\newtheoremstyle{exercise}% name
  {3pt}%      Space above
  {6pt}%      Space below
  {}%         Body font
  {}%         Indent amount (empty = no indent, \parindent = para indent)
  {\bfseries}% Thm head font
  {:}%        Punctuation after thm head
  { }%     Space after thm head: " " = normal interword space;
   {}%         Thm head spec (can be left empty, meaning `normal')
\theoremstyle{exercise}
\newtheorem{exercise}[theorem]{Exercise}
\newtheoremstyle{exercises}% name
  {3pt}%      Space above
  {6pt}%      Space below
  {}%         Body font
  {}%         Indent amount (empty = no indent, \parindent = para indent)
  {\bfseries}% Thm head font
  {:}%        Punctuation after thm head
  {\newline}%     Space after thm head: " " = normal interword space;
   {}%         Thm head spec (can be left empty, meaning `normal')

\theoremstyle{exercise}
\newtheorem{exercises}[theorem]{Exercises}

\def\boxit#1{\vbox{\hrule height1pt\hbox{\vrule width1pt\kern3pt
  \vbox{\kern3pt#1\kern3pt}\kern3pt\vrule width1pt}\hrule height1pt}}

\def\bv{\bold v}

\def\BC{\mathbb C}

\def\tdim{{\rm dim}}

\def\hd{,...,}
\def\ww{\wedge}

\def\inv{{}^{-1}}

\def\11{\mathbf 1}

\def\a{\alpha}

\def\b{\beta}
\def\g{\gamma}
\def\s{\sigma}

\def\ot{{\mathord{ \otimes } }}

\def\otc{{\mathord{\otimes\cdots\otimes}\;}}

\def\ra{{\mathord{\;\rightarrow\;}}}

\def\dim{{\rm dim}\;}

\def\s{\sigma}
\def\t{\tau}

\def\a{\alpha}
\def\b{\beta}

\def\g{\gamma}

\def\FS{\mathfrak  S}

\def\BC{\mathbb  C}

\def\hd{, \hdots ,}

\def\inv{{}^{-1}}

\def\ur{\underline {\bold R}}

\def\ra{\rightarrow}

\def\tdim{\operatorname{dim}}

\def\ww{\wedge}

\def\bbb{{\bold{b}}}

\def\be{\begin{equation}}
\def\ene{\end{equation}}
\def\aaa{{\bold {a}}}
\def\bbb{{\bold {b}}}
\def\ccc{{\bold {c}}}
\def\tsgn{{\rm{sgn}}}

\def\IC{\mathbb{C}}

\def\aaa{{\bold a}}\def\bbb{{\bold b}}\def\ccc{{\bold c}}

\newcommand{\red}[1]{{#1}}
\newcommand{\blue}[1]{{#1}}
\newcommand{\green}[1]{{#1}}
\newcommand{\magenta}[1]{{#1}}
\newcommand{\cyan}[1]{{{#1}}}
\newcommand{\strike}[1]{}
\newcommand{\todo}[2]{#1}

\def\sL{{\mathcal L}}
\def\sK{{\mathcal K}}

\let\brright=)

\newcommand{\bC}{{\mathbb C}}
\newcommand{\bP}{{\mathbb P}}
\newcommand{\sH}{{\mathcal H}}
\newcommand{\Span}{{\rm span}}

\begin{document}
\raggedbottom

%\title[Computer aided methods for lower bounds on the border rank]{Computer aided methods for \\ lower bounds on the border rank}
\title[\blue{Equations for lower bounds \magenta{on border} rank}]{\blue{Equations for\\lower bounds \magenta{on border} rank}}
\author{Jonathan D. Hauenstein, Christian Ikenmeyer, and  J.M. Landsberg}
%\date{April 2010}
 \begin{abstract}
We present new methods for determining polynomials in the ideal of
the variety of bilinear maps of border rank at most $r$.
We apply these methods to several cases including the
case $r = 6$ in the space of bilinear maps $\BC^4\times\BC^4\ra\BC^4$.
This space of bilinear maps includes the matrix multiplication operator $M_2$ for
two by two matrices.  We show these newly obtained polynomials
do not vanish on the matrix multiplication operator $M_2$, which gives  a new proof that the border rank of the multiplication
of $2\times 2$ matrices is seven.  Other examples are considered
along with an explanation of how to implement the methods.
\end{abstract}
\thanks{Research of Hauenstein supported in part by AFOSR grant FA8650-13-1-7317 and NSF grant DMS-1262428.
Research of Ikenmeyer supported in part by DFG grant BU 1371/3-2.
Research of Landsberg supported in part by NSF grant DMS-1006353.}
\email{hauenstein@ncsu.edu, ciken@math.upb.de, jml@math.tamu.edu}
\maketitle

\subsection*{Acknowledgements}
We thank Peter B\"urgisser for important discussions and suggestions,
\green{and the anonymous reviewer for many helpful comments}.

\section{Introduction}

Lower bounds in complexity theory are considered difficult to obtain.
We describe a new method for obtaining lower bounds on the {\it border rank}
which is based on a new way to find polynomials that vanish on
bilinear maps $T:\BC^{\aaa}\times \BC^{\bbb}\ra \BC^{\ccc}$ of low border rank.

\subsection{Rank and border rank}

Let $\BC^{\aaa *}:=\{ f: \BC^{\aaa}\ra \BC\mid f {\rm \ is \ linear}\}$
denote the dual vector space to $\BC^{\aaa}$.  That is, if an element of $\BC^{\aaa}$
is represented by a column vector of height $\aaa$,
then $\BC^{\aaa*}$ corresponds to row vectors, and the evaluation is just
row-column matrix multiplication.  A bilinear map $T\colon\BC^{\aaa}\times \BC^{\bbb}\ra \BC^{\ccc}$ has
{\it rank one} if there exist $\a\in \BC^{\aaa *}$, $\b\in \BC^{\bbb*}$,  and $c\in \BC^{\ccc}$ such that
$T(a,b)=\a(a)\b(b)c$.
The rank one bilinear maps are in some sense the simplest bilinear maps, and
$T$ is said to have {\it rank} $r$ if $r$
is the minimum number of rank one bilinear maps which sum to $T$.
This $r$ is sometimes called the {\it tensor rank} of $T$.
If one views multiplication by constants as a ``free'' operation,
then the rank differs at most by a factor of two
from the minimal number of multiplications of variables that
is needed to compute $T$, see \cite[Ch.~14]{BCS:97} for more information. %ciken 4-29

Since the set of all bilinear maps $\BC^{\aaa}\times \BC^{\bbb}\ra\BC^{\ccc}$ is a vector space of dimension $\aaa\bbb\ccc$,
it is natural to talk about polynomials on the space of bilinear maps $\BC^{\aaa}\times \BC^{\bbb}\ra \BC^{\ccc}$.
Unfortunately, one cannot test directly for the tensor rank by the vanishing of polynomials,
since the common zero locus of the set of all polynomials vanishing on the set of bilinear maps of rank at most $r$
is, typically, larger than the set of bilinear maps of rank at most $r$.
This may be described precisely using the language of algebraic geometry: for the purposes of this article,
we define an  {\it algebraic variety} (or simply a {\it variety})  to be the common zero locus of a collection of polynomials that is
{\it irreducible}, in the sense that it cannot be written as a union of two zero loci.

A (proper) {\it Zariski closed subset}
of a variety $X$ is the common zero locus of a collection of polynomials restricted to $X$, and a {\it Zariski open subset}
is the complement of a Zariski closed set.
\blue{T}he  {\it border rank} of a tensor  $T$ is defined to be the smallest
 $r$ such that  all polynomials vanishing on the set of bilinear maps of rank
at most $r$ also vanish at $T$, and one writes $\ur(T)= r$.
In this case, $T$ is arbitrarily close, in any \red{reasonable} measure, to a bilinear map
of rank   $r$ (including the possibility that the rank of $T$ is $r$).
We let $\s_{r;\aaa,\bbb,\ccc}\blue{\subset \BC^{\aaa}\ot \BC^{\bbb}\ot \BC^{\ccc}}$ denote the
 set of  bilinear maps of border rank at most $r$. It is \blue{the  algebraic variety  formed from the zero set of %jml 4-26
all the polynomials having the set of bilinear maps of rank at most $r$ in their zero set.}
When $\aaa,\bbb,\ccc$ are understood from the context, we simply write $\s_r$.
\blue{The set of bilinear maps of rank $r$ is a   Zariski open subset
of the algebraic variety  $\s_{r;\aaa,\bbb,\ccc}$.  The set is open because   the set of
bilinear maps of border rank less than $r$ is a closed subset of $\BC^{\aaa}\ot \BC^{\bbb}\ot \BC^{\ccc}$,
and  the subset   of $\s_{r;\aaa,\bbb,\ccc}$ of bilinear maps of rank greater than $r$ is closed in $\s_{r;\aaa,\bbb,\ccc}$.}

\subsection{Results}\label{Sec:Results}

We introduce a new technique based on numerical algebraic geometry and interpolation
that finds, with high probability, where equations which vanish on the variety of bilinear maps of border rank
at most $r$ can be found.  Once one knows where to look, we can use methods which
began in \cite{LMsec} and were refined in \cite{batesoeding,BItensor}
to find the actual equations \red{and \blue{rigorously prove they vanish on $\s_{r;\aaa,\bbb,\ccc}$}}.
\strike{The vanishing of these equations can be sometimes proved rigorously and other times proven with extremely high probability depending on the application.}
With these equations, \green{one can then show that the border rank of a given tensor $T$ is greater than $r$ if $T$ does not satisfy these equations}.
Of special interest in this paper will be border rank of the matrix multiplication tensor
\[
 M_2 := \sum_{i,j,k = 1}^2 e_{i,j} \otimes e_{j,k} \otimes e_{k,i} \in \bC^4 \otimes \bC^4 \otimes \bC^4,
\]
where $(e_{i,j})$ is the standard basis of $\bC^{2\times 2} = \bC^4$. %jml 4-26

\red{
It is known since Strassen's fundamental 1969 breakthrough \cite{str:69} that $\ur(M_2) \leq 7$.
Our main result is \magenta{a new proof of} the lower bound}
\begin{equation}\label{Eq:BorderRank7}
\ur(M_2) \geq 7,
\end{equation}

\smallskip

\noindent \magenta{which was originally proven in 2005 by Landsberg, see} \cite{Lmatrixarxiv},
\blue{where very different methods were used (see Section~\ref{Sec:ReviewProof})}.
The proof outline is as follows.
We start by \blue{proving, with the aid of a computer,} that no nonconstant polynomial of degree less than $19$ vanishes on $\s_{6;4,4,4}$.
\green{In the course of this computation, we compute the necessary data to perform the membership test of
\cite{HS13} which numerically shows (i.e., this shows with extremely high probability) that~\eqref{Eq:BorderRank7} holds.
Additionally,} the same data gives strong evidence that there is a $64$\blue{-d}imensional space of
degree $19$ equations that vanish on $\s_{6;4,4,4}$.
The only $64$\blue{-d}imensional representation of $GL_4$ in $\IC[\blue{\BC^{4}\ot \BC^{4}\ot \BC^{4}}]_{19}$ is
of type $((5,5,5,4),(5,5,5,4),(5,5,5,4))$.
By a randomized procedure, we then construct a basis for the \magenta{31}-dimensional
highest weight vector space of weight $((5,5,5,4),(5,5,5,4),(5,5,5,4))$ in $\IC[\blue{\BC^{4}\ot \BC^{4}\ot \BC^{4}}]_{19}$.
\blue{We show using numerical methods that the} restriction of this \green{31}-dimensional vector space
to functions defined on $\s_{6;4,4,4}$ has a 1-dimensional kerne\blue{l}.

\green{Since} the highest weight \blue{s}pace of \blue{the} weight $((5,5,5,5),(5,5,5,5),(5,5,5,5))$
in $\IC[\blue{\BC^{4}\ot \BC^{4}\ot \BC^{4}}]_{20}$ is only 4-dimensional,
\green{we focus on this space to develop a rigorous proof of~\eqref{Eq:BorderRank7}.
In particular, the restriction to functions defined on $\s_{6;4,4,4}$ also
has a 1-dimensional kernel.  This corresponds to a degree $20$ polynomial that vanishes on $\s_{6;4,4,4}$
which does not vanish at $M_2$ thereby completing the proof.}

\blue{We remark that while there is a large subspace of $\IC[\blue{\BC^{4}\ot \BC^{4}\ot \BC^{4}}]_{20}$
that vanishes on $\s_{6;4,4,4}$,   the polynomial we work with is distinguished in that it is the only one
that is unchanged (up to scale) by changes of bases in each of the $\BC^4$'s.}

To make this approach computationally feasible, \green{each
polynomial is represented by a pair of permutations, see Section~\ref{subsec:firstalgo}.
These permutations provide all the information needed to evaluate the corresponding polynomial,
but, unfortunately, \magenta{this} means that we are unable to obtain additional information, such as
the number of terms.}

The technique we present can be applied to other {\em implicitization problems}.
That is, we want to consider a variety
\begin{equation}\label{eq:ImageX}
X := \overline{g(Y)}
\end{equation}
where $Y$ is (possibly a Zariski open set of) a variety  and $g$ is a system of
rational functions defined on $Y$.
In the bilinear case, the tensors of rank at most $r$ is a dense subset
of the algebraic set of tensors of border rank at most $r$
where each tensor of rank at most $r$ can be written as a sum of
$r$ tensors of rank at most one.  In this case, $Y$ is simply the
product of $r$ copies of the variety of tensors of rank at most one
and $g$ is the linear map corresponding to taking the sum.
Another specific application arising in physics is the analysis of vacuum moduli space
in (supersymmetric) field theories \cite{HHM12} which arise
as the closure of the image under a polynomial map of an algebraic~set.

\red{Besides our main result, \blue{we proved, using numerical methods} the following:}
\begin{itemize}
\item The degree of $\s_{6;4,4,4}$ is 15,456.
\item \green{The degree of the codimension three variety $\s_{15;4,8,9}$ is at~least~83,000
    and no nonconstant polynomial of degree $\leq 45$ vanishes on $\s_{15;4,8,9}$.}
\item The degree of the hypersurface $\s_{18;7,7,7}$ is at~least~187,000.
\item The degree of the codimension six variety $\s_{6;3,4,6}$ is 206,472 and \green{no nonconstant polynomial of degree $\leq 14$ vanishes on $\s_{6;3,4,6}$.}
\item The degree of the codimension three variety $\s_{7;4,4,5}$ is 44,000 and \green{no nonconstant polynomial of degree $\leq 56$ vanishes on $\s_{7;4,4,5}$.}
\item The degree of the hypersurface $\s_{8;3,5,7}$ is $105$.
\end{itemize}

\green{The varieties $\s_{15;4,8,9}$ and $\s_{18;7,7,7}$ have applications to $3\times 3$ matrix multiplication.
Information about polynomials in their ideals could help to more precisely determine $\ur(M_3)$, 
with the current known bounds being $15 \leq \ur(M_3) \leq 21$\magenta{, see \cite{MR623057} and \cite{LOsecbnd}}.
The other varieties are presented since they are the margin of what is currently feasible.
Results regarding the ideal being empty are potentially useful for finding further equations
since they provide a starting point for~such~an~endeavor.}  

\subsection{Other methods for finding equations}

Very little is known about the equations of $\s_r$ in general.
One can reduce to the case of $\aaa=\bbb=\ccc=r$ via a process called {\it inheritance}.
Additionally, there is a systematic way to determine the equations in any given degree using {\it multi-prolongation}.
For a discussion on inheritance and multi-prolongation, see \cite[\S 3.7]{MR2865915}.
Even though multi-prolongation is systematic, it is very difficult to utilize except in very small cases.
Most known equations have been found by reducing multi-linear algebra to linear algebra.
See \cite{Lhighbranktensor,LOsecbnd} for the most recent equations that go up to $\s_{2m-2;m,m,m}$.
Some information about the ideal of  $\s_{r;r,r,r}$ can be found using representation theory (via the algebraic Peter-Weyl Theorem)
as this case is an orbit closure, see \cite{BIrank} for an exposition. By inheritance one could deduce the $\s_{r;m,n,p}$ case
for any $m,n,p$ from
the $\s_{r;r,r,r}$ case.

\subsection{Polynomials on vector spaces}

We write $I(\s_r)$ for the set of all polynomials vanishing on $\s_r$, which forms an {\em ideal}.
% (The set of all polynomials vanishing on $\s_r$ is
%indeed an ideal as if $P,Q\in I(\s_r)$, then $P+Q\in I(\s_r)$ and $PF\in I(\s_r)$ for any polynomial $F$.)
Since $\s_r$ is invariant under re-scaling, we may restrict our attention to
homogeneous polynomials since, in this case, a polynomial will be in the ideal
if and only if all of its homogeneous components are in the ideal.

Let $V$ be a vector space. A subset $X\subset V$ is called an {\em algebraic set}
if it is the common zero locus of a collection of polynomials on~$V$.
Recall that we say that an irreducible algebraic set is a {\it variety}. %ciken 4-26
If $X\subset V$ is a variety that is invariant under re-scaling,
let $S^d V^*$ be the space of homogeneous polynomials of
degree $d$ on $V$ and $I_d(X)\subset S^d V^*$ be
the component of the ideal of~$X$ in degree~$d$.

Roughly speaking (see Section~\ref{sec:Alg} for more details),
our technique for studying the equations that vanish on a variety $X$ of
positive dimension is by applying \cyan{numerical algebraic geometry %and machine learning
techniques} to finite subsets of $X$ which lie in a common linear space.
That is, we aim to study finite subsets of algebraic sets of the
form $Y = X\cap \sL\subset \sL$ where $\sL$ is a general linear space of
codimension at most $\dim X$.  If the codimension of $\sL$ is $\dim X$,
then $Y$ consists of $\deg X$ points.
If the codimension of $\sL$ is strictly less than $\dim X$, then $Y$ is also
a variety with the same degree as $X$.
Moreover, if one considers $X\subset V$ and $Y\subset\sL$,
and defines $d_X$ and $d_Y$ to be the minimal degree of the nonzero polynomials in
$I(X)$ and $I(Y)$, respectively, then $d_X \geq d_Y$.
In particular, $\dim I_d(Y)\geq\dim I_d(X)$ for any $d\leq d_X$
with similar bounds for all $d\geq0$ that can be developed from the corresponding Hilbert functions.
Once we have inferred information about polynomials in $I(Y)$,
we use representation theory to identify which modules could appear.
Finally, sample vectors from these modules are used to test if the entire module is in the ideal $I(X)$ or not.

\section{Deciding where to go hunting}\label{sec:Alg}

The basic idea of our algorithm is to combine the ability of numerical
algebraic geometry to compute points on certain subsets of a variety
with \green{interpolation} to obtain information about this subset from
the computed points.  We first describe needed concepts from numerical algebraic geometry
and then \green{a brief discussion regarding interpolation}.

At a basic level, the algorithms of numerical algebraic geometry
(see \cite{SW05} for general background information)
perform numerical computations on varieties where each variety is represented
by a data structure called a {\em witness set}.
Let $f$ be a polynomial system.  The common zero locus of $f$ is an algebraic set
that can be decomposed uniquely into finitely many varieties,
none of which is contained in the union of the others.
If $X$ is one of these varieties, called an {\em irreducible component}
of the zero locus of $f$, then a witness set for $X$ is the triple $\{f,L,W\}$
where the zero set of $L$ defines a general linear subspace of
codimension equal to the dimension of $X$ and $W$ is the intersection of $X$ with this linear subspace defined by $L$.
Given one point in $W$, arbitrarily many points on $X$ can be computed in a process called {\em sampling}.
In numerical terms, computing a point ``on'' a variety means that we have
both a numerical approximation of the point along with an
algorithm that can be used to approximate the point~to~arbitrary~accuracy.

This witness set description is not useful for the problems
at hand since, for each of the varieties $X$ under consideration, we do not
assume that we have access to a polynomial system $f$ let alone {\em any} nonzero polynomials which
vanish on $X$.  However, we do assume that we
have a description of $X$ in the form \eqref{eq:ImageX}.
In fact, by adding variables and clearing denominators, we can assume
that $X := \overline{\pi(Z)}$ where $\pi$ is a projection map
and $Z$ is an irreducible component of the zero locus for some polynomial system $F$.
This is demonstrated in the following~simple~example.

\begin{example}
The set $X := \{(x,y)\in\bC^2~|~x^2 + y^2 = 1\}$ is equal to $\overline{g(Y)}$ where
$$g(t) = \left(\frac{1-t^2}{1+t^2},\frac{2t}{1+t^2}\right) \hbox{~and~} Y := \bC\setminus\{\pm i\}.$$
We also have $X = \overline{\pi(Z)}$ where $\pi(x,y,t) = (x,y)$ and $Z$ is the zero locus (which is irreducible) of
$$F(x,y,t) = \left[\begin{array}{c} (1+t^2)x - (1-t^2) \\ (1+t^2)y - 2t \end{array}\right].$$
\end{example}

With this setup, we utilize a {\em pseudowitness set} \cite{HS10,HS13} for $X = \overline{\pi(Z)}$ 
which is the quadruple $\{F,\pi,L,W\}$ where $L$ defines a linear subspace of codimension equal
to the dimension of $Z$ and $W$ is the intersection of $Z$ with this linear subspace
defined by $L$.  In this case, the linear polynomials $L$ are constructed so that it
has exactly $\dim X$ general linear polynomials in the image space of $\pi$,
i.e., intersect $X$ in $\deg X$ many points, while the remaining linear
polynomials are general.  In particular, $\pi(W)$ consists of exactly $\deg X$ distinct points.
As with traditional witness sets, one can sample and perform membership
tests on $X$ \cite{HS13}.

The key here is that once a single sufficiently general point is known
on $X$, other points on $X$ can be computed as well.  In fact, these
other points can be forced to live in a fixed general linear subspace
of codimension at most $\dim X$ thereby simplifying the future computations since one can work
intrinsically on this linear subspace.  If the intersection of
the linear subspace and $X$ is positive dimensional, then it is also a variety
and arbitrarily many points can be sampled from this variety.
If the intersection is zero-dimensional, it consists of exactly $\deg X$
points which, after computing one, random monodromy loops \cite{Monodromy}
could be used to try to compute the other points.
The trace test \cite{Trace} provides a stopping criterion for deciding
when exactly $\deg X$ points have been computed.

Clearly, any polynomial which vanishes on $X$ must also vanish on a
finite subset of $X$.  Although we will not delve too deep into the theory here,
one can recover the invariants of $X$ from a general linear subspace section of $X$
when $X$ is an {\em arithmetically Cohen-Macaulay} scheme (see \cite[Chap. 1]{Migliore}).
Nonetheless, since our current focus is on developing a list of potential
places of where to look to focus further representation theoretic computations,
we can consider all varieties and not just the arithmetically Cohen-Macaulay ones.
Of course, this is at the expense of bounds rather than equality as demonstrated in the following example.

\begin{example}\label{ex:SimpleEx}
Consider the following   varieties in $\bP^3$: %jml-4/26
$$X_1 := \{(s^3,s^2 t,st^2,t^3)~|~(s,t)\in\bP^1\} \hbox{~and~} X_2 := \{(s^4,s^3 t,st^3,t^4)~|~(s,t)\in\bP^1\}.$$
It is easy to verify that
\begin{itemize}
\item $\dim X_1 = 1$, $\deg X_1 = 3$, and $I(X_1)$ is generated by three quadratics;
\item $\dim X_2 = 1$, $\deg X_2 = 4$, and $I(X_2)$ is generated by a quadratic and three cubics.
\end{itemize}
Let $Y_i = X_i \cap \sH$ be the set of $\deg X_i$ points where $\sH$ is the hyperplane defined by
the vanishing of $\ell(x) = x_0 + 2x_1 + 3x_2 + 5x_3$.  If we consider $Y_i\subset\sH$, then
\begin{itemize}
\item $I(Y_1)$ is generated by three quadratics;
\item $I(Y_2)$ is generated by two quadratics.
\end{itemize}
To summarize, $X_1$ is the twisted cubic curve in $\bP^3$
which is arithmetically Cohen-Macaulay so that, for example,
the dimension of $I_d(X_1)$ can be determined from $I_d(Y_1)$.
However, $X_2$ is not arithmetically Cohen-Macaulay which, in this case,
can be observed since $2 = \dim I_2(Y_2) > \dim I_2(X_2) = 1$.
Even though one should only expect $d_{X_2} \geq d_{Y_2}$, we do have equality
in this case, namely $d_{X_2} = d_{Y_2} = 2$.
\end{example}

\green{
Once we have decided on our first finite set to consider,
the next task is {\em polynomial interpolation}, that is, 
to compute polynomials that vanish on this finite set.
Given a basis for the finite-dimensional space of polynomials under 
consideration, polynomial interpolation reduces to computing null vectors
of a (potentially very large) matrix.
From a numerical standpoint, as the degrees of the polynomials under consideration 
increase, preconditioning becomes essential to perform reliable computations.  
For our computations, we use the approach of \cite{GHPS}.}

\green{Each computation} provides some restrictions on which
polynomials can be in $I(X)$.  Nevertheless, we also consider what happens when we add new points to
our finite set.  For a particular degree, there are two possible choices: either the originally
computed polynomials will vanish at the new points
or the dimension of the set of polynomials that vanish at all the points will decrease.
In the former case, we can then move on to searching for higher degree polynomials not
generated by these polynomials.  In the latter case, we continue adding new points.
If no polynomials of a particular degree, say $d$, vanish on some finite set,
then we know that $\dim I_d(X) = 0$ and $d_X > d$.  Thus, we try again by considering
polynomials of degree $d+1$.

Variations of this approach can be to consider sampling points from
the intersection of $X$ with linear spaces of increasing dimension
to see how the dimension of the vanishing polynomials change as less
restrictions are placed on the sample points.  The key in the end is to
control the growth of the dimension of the space of polynomials under
consideration since this can become unwieldy quickly.  In particular,
this method is practical for varieties $X$ of low codimension
since we can work implicitly on linear spaces of low dimension.

When the codimension is one, $X$ is a hypersurface so that the degree of $X$
is equal to the degree of the polynomial defining $X$.
In this case, one can simply compute a pseudowitness set
to compute its degree rather then use this interpolation based approach.
For example, such an approach was used in \cite{BHORS} for
computing the degree of implicitly defined hypersurfaces,
which arise as the algebraic boundaries of Hilbert's sums of squares cones
of degree 38,475 and 83,200.

\begin{example}
In Example~\ref{ex:SimpleEx}, we considered finite sets obtained by intersecting the curves
with a particular hyperplane.  We now use this information to limit our focus when we add other points
to our finite set.  In four variables, there is a ten-dimensional space of homogeneous polynomials of degree $2$,
but with our previously computed information, this has already been reduced to a seven and six dimensional space
for $X_1$ and $X_2$, respectively.  More specifically, the four dimensional space arising from the linear
polynomial $\ell(x)$ along with the three and two dimensional spaces, respectively, from $I_2(Y_1)$ and $I_2(Y_2)$, namely
\begin{itemize}
\item $I_2(X_1)\subset\Span\left\{\begin{array}{l}
x_0\ell(x),x_1\ell(x),x_1 x_2+2x_1 x_3+3x_2 x_3+5x_3^2,\\
x_2\ell(x),x_3\ell(x),x_2^2-x_1 x_3,x_1^2-x_1 x_3-x_2 x_3-10 x_3^2\end{array}\right\};$
\item $I_2(X_2)\subset\Span\left\{\begin{array}{l}
x_0\ell(x),x_1\ell(x),x_1 x_2+2x_1 x_3+3x_2 x_3+5x_3^2,\\
x_2\ell(x),x_3\ell(x),x_1^2-x_2^2+11x_1 x_3+2x_2 x_3+20x_3^2\end{array}\right\}.$
\end{itemize}
By selecting additional random points, one indeed finds $\dim I_2(X_1) = 3$ and $\dim I_2(X_2) = 1$.
\end{example}

This procedure develops ideas on where the degrees $d_j$ in which
generators of the ideal appear.  
\green{The next section summarizes 
the numerical evidence for the results presented in Section~\ref{Sec:Results}.
From this data, the next step} is to conclusively determine the linear subspace
of the space of polynomials of degrees $d_j$ that are in the ideal. 
For this, one uses representation theory as we describe in Section~\ref{Sec:PolyBilinear}.

\section{Review of numerical results}\label{Sec:ReviewNumResults}

\green{We summarize the six varieties presented in Section~\ref{Sec:Results}.  In all these cases, the codimension 
of the variety is the expected codimension, namely $\codim \s_{r;\aaa,\bbb,\ccc} = \aaa\bbb\ccc - r(\aaa + \bbb + \ccc - 2)$.
The points on the varieties were computed using Bertini \cite{Bertini} with the linear algebra computations performed using Matlab.
}

\subsection{$\s_{6;4,4,4}$}

\green{
As discussed above, this codimension four variety provides information about $\ur(M_2)$, that is, 
showing $M_2\notin\s_{6;4,4,4}\subset\bP^{63}$ shows $\ur(M_2) = 7$.  
We first fix a random linear space $\sL\subset\bP^{63}$ of dimension $4$ and consider 
the finite set $W := \s_{6;4,4,4}\cap\sL$.  The first objective is to compute points in
$W$, with a goal of computing every point in $W$.  To this end, we first computed one point in $W$
as follows.  One first picks a random point $x^*\in\s_{6;4,4,4}$, which is trivial since a dense subset is 
$\s_{6;4,4,4}$ is parameterizable.  Let $L$ be a system of $59$ linear forms so that $\sL$ is the zero locus of $L$
and $\sL_{t,x^*}$ be the zero locus of $L(x) - t\cdot L(x^*)$.  Since $x^*\in \s_{6;4,4,4}\cap\sL_{1,x^*}$, 
a point in $W$ is the endpoint of the path defined by $\s_{6;4,4,4}\cap\sL_{t,x^*}$ at $t = 0$ starting
from $x^*$ at $t = 1$.  

Even though the above process could be repeated for different $x^*$ to compute points in $W$,
we instead used monodromy loops \cite{Monodromy} for generating more points in $W$.  
After performing $21$ loops, the number of points in $W$ that had computed stabilized at 15,456.  
The trace test \cite{Trace} shows that 15,456 is indeed the degree of $\s_{6;4,4,4}$ thereby
showing we had indeed computed $W$.  

From $W$, we performed two computations.  The first was the membership test of \cite{HS13}
for deciding if $M_2\in\s_{6;4,4,4}$, which requires tracking 15,456 homotopy paths that start at the points of $W$.
In this case, each of these 15,456 paths converged to points in $\s_{6;4,4,4}$ distinct from $M_2$
providing a numerical proof that $M_2\notin\s_{6;4,4,4}$.  
The second was to compute the minimal degree of nonzero polynomials vanishing on $W\subset\sL$.
This sequence of polynomial interpolation problems showed that no nonconstant polynomials 
of degree $\leq 18$ vanished on $W$ and hence $\s_{6;4,4,4}$.  
The $15456\times 8855$ matrix resulting from polynomial interpolation of
homogeneous forms of degree $19$ in $5$ variables using the approach of \cite{GHPS}
has a 64-dimensional null space.
Thus, the minimal degree of nonzero polynomials vanishing on $W\subset\sL$ is $19$.  

The next objective was to verify the minimal degree of nonzero polynomials
vanishing on the curve $C := \s_{6;4,4,4}\cap\sK \subset\sK$
for a fixed random linear space $\sK\subset\bP^{63}$ of dimension $5$
was also $19$.  We used 50,000 points on $C$ and
the $50000\times 42504$ matrix resulting from polynomial interpolation of
homogeneous forms of degree $19$ in $6$ variables using the approach of \cite{GHPS}
also has a 64-dimensional null space.
With this agreement, we proceeded to use representation theory,
described in Section~\ref{Sec:PolyBilinear}, to understand these polynomials and 
prove that $M_2$ is indeed not contained in $\s_{6;4,4,4}$.  
}

\subsection{$\s_{15;4,8,9}$}

\green{
For this codimension three variety, we followed a similar computation
as above for computing $83,000$ points in $W := \s_{15;4,8,9}\cap\sL$
where $\sL\subset\bP^{287}$ is a random linear space
of dimension $3$.  By using polynomial interpolation on these points, 
we are able to show that no nonconstant polynomial of degree $\leq 45$ 
vanishes on $\s_{15;4,8,9}$.
}

\subsection{$\s_{18;7,7,7}$}

\green{
For this hypersurface, we computed $187,000$ points in $W := \s_{18;7,7,7}\cap\sL$
where $\sL\subset\bP^{342}$ is a random line.  This shows
that $187,000$ is a lower bound on the degree of $\s_{18;7,7,7}$ and the degree of the polynomial which defines it.
}

\subsection{$\s_{6;3,4,6}$}

\green{
For this codimension six variety, we followed a similar computation
as above for computing $W := \s_{6;3,4,6}\cap\sL$
where $\sL\subset\bP^{71}$ is a random linear space
of dimension $6$.  In this case, the trace test shows that the set of
206,472 points computed by monodromy does indeed equal $W$.
Polynomial interpolation showed that no nonconstant polynomial
of degree $\leq 14$ vanished on $\s_{6;3,4,6}$.  We stopped at degree $14$ due
to memory limitations of the numerical linear algebra routines.  However, even though
we were unable to compute the minimal degree of nonconstant polynomials vanishing
on $W\subset\sL$, we note that $W$ with \cite{HS13} can be used
to decide membership in $\s_{6;3,4,6}$.
}

\subsection{$\s_{7;4,4,5}$}

\green{
For this codimension three variety, we followed a similar computation
as above for computing $W := \s_{7;4,4,5}\cap\sL$
where $\sL\subset\bP^{79}$ is a random linear space
of dimension $3$.  In this case, the trace test shows that the set of
44,000 points computed by monodromy does indeed equal $W$.
Polynomial interpolation showed that no nonconstant polynomial
of degree $\leq 56$ vanished on $\s_{7;4,4,5}$.  We stopped at degree $56$ due
to additional conditioning problems arising from the numerical linear algebra routines.  
As with the $\s_{6,3,4,6}$ case, $W$ with \cite{HS13} still can be used
to decide membership in $\s_{7;4,4,5}$.
}

\subsection{$\s_{8;3,5,7}$}

\green{
For this hypersurface, the trace showed that the set of 105 points computed by
monodromy is equal to $W := \s_{8;3,5,7}\cap\sL$
where $\sL\subset\bP^{104}$ is a random line.  
In particular, this shows that there is a degree 105 polynomial vanishing on $\s_{8;3,5,7}$.
}

\section{Polynomials on the space of bilinear maps}\label{Sec:PolyBilinear}

\subsection{Tensors}

In order to explain the polynomials it will be useful to work more invariantly, so instead of
$\BC^{\aaa},\BC^{\bbb}$ etc., we write $A,B$ etc.   for complex vector spaces of dimensions $\aaa,\bbb$ etc..
It will also be useful to introduce the language of {\it tensors}. A  bilinear map $A^*\times B^*\ra C$ may also be viewed
as a tri-linear map $A^*\times B^*\times C^*\ra \BC$, as well as in numerous other ways. To avoid prejudicing ourselves,
we simply write
 $T\in A\ot B\ot C$ for any of these manifestations and call $T$ a {\it tensor}. Just as we may view a linear map as a matrix
after fixing bases, such $T$ may be viewed as a three-dimensional matrix after fixing bases. Note that
$A\ot B\ot C$, the set of all such tensors, is a vector space of dimension $\aaa\bbb\ccc$.
More generally, given vector spaces $A_1\hd A_k$, one can define the space of tensors $A_1\otc A_k$.
There is a natural map $A_1\otc A_k\times B_1\otc B_l\ra A_1\otc A_k\ot B_1\otc B_l$, $(f,g)\mapsto f\ot g$,  where
$f\ot g(\a_1\hd a_k,\b_1\hd \b_l):=f(\a_1\hd a_k)g(\b_1\hd \b_l)$.

\subsection{Remarks on the theory}

We briefly review the representation theory underlying  in the algorithm. For more details, see
\cite[Chap. 6]{MR2865915}.
 Let $S^d(A \ot B \ot C )^*$ denote the
vector space of homogeneous polynomials of degree $d$ on $A\ot B\ot C$.
The variety $\s_{r;\aaa,\bbb,\ccc}$ is mapped to itself under changes of bases in each of the
vector spaces and thus if we have one equation, we can obtain many more
by changing bases. That is, let $GL(A)$ denote the set of invertible linear maps $A\ra A$ and similarly for $B,C$.
The group $G:=GL(A)\times GL(B)\times GL(C)$ acts on
$A\ot B\ot C$ by $(g_A,g_B,g_C)\cdot (\sum_i a_i\ot b_i\ot c_i)=
\sum_i g_A a_i\ot g_B b_i\ot g_C c_i$, and $GL(V)$ acts on $S^d V^*$ by
$g\cdot P(x)=P(g\inv \cdot x)$. Letting $V=A\ot B\ot C$ and noting
$G\subset GL(V)$, we have a $G$-action on  $S^d(A \ot B \ot C )^*$. If $P\in I(\s_r)$, then
$g\cdot P\in I(\s_r)$ for all $g\in G$.  Since ideals
are in particular vector spaces,  the linear span of the orbit of $P$  in $S^d(A \ot B \ot C )^*$
will be in $I(\s_r)$.

We will use the action of the group $G$ to organize our calculations. A group $G$ is said to {\it act}  on
a vector space $V$ if there is a group homomorphism $\rho: G\ra GL(V)$. Then $V$ is called a {\it $G$-module}. The $G$-module
$V$ is said to be {\it irreducible} if there is no nontrivial subspace of $V$ invariant under the action of $G$.
The irreducible polynomial $GL(V)$ modules are indexed by partitions $\pi=(p_1\hd p_{\bv})$,
where $p_1\geq\cdots\geq  p_{\bv}\geq 0$. We write $|\pi|=p_1+\cdots +p_{\bv}$, and we say $\pi$ is a partition of $|\pi|$.
Let $S_{\pi}V$ denote the corresponding irreducible $GL(V)$-module. It occurs in $V^{ \ot |\pi |}$ and no
other tensor power, however not uniquely - there is a vector space's worth of realizations
except in the cases $\pi=(d)$ or $\pi=(1\hd 1)$.
 The irreducible $GL(A)\times GL(B)\times GL(C)$-modules are
all of the form $V_A\ot V_B\ot V_C$ where $V_A$ is an irreducible $GL(A)$-module etc..
 For $G=GL(A)\times GL(B)\times GL(C)$, every $G$-module decomposes into a direct sum of irreducible
submodules. This decomposition is not unique in general, but the {\it isotypic} decomposition, where
all isomorphic modules are grouped together, is.

We are interested in the homogeneous  polynomials of degree say $d$  on $A\ot B\ot C$,
denoted $S^d(A\ot B\ot C)^*$. Via {\it polarization},  a polynomial may be considered as a symmetric
tensor so $S^d(A\ot B\ot C)^*\subset (A\ot B\ot C)^{*\ot d}\simeq A^{* \ot d}\ot   B^{* \ot d}\ot  C^{* \ot d}$.
Thus, the isomorphism types of irreducible $G$-modules in $S^d(A\ot B\ot C)^*$ are described by triples
$(\pi,\mu,\nu)$ of partitions of $d$ whose number of parts $\ell(\pi)\leq \aaa$ etc.. Let
$k_{\pi,\mu,\nu}$ denote the multiplicity of $S_{\pi}A^*\ot S_{\mu}B^*\ot S_{\nu}C^*$ in $S^d(A\ot B\ot C)^*$,
that is, the dimension of the space of realizations of $S_{\pi}A^*\ot S_{\mu}B^*\ot S_{\nu}C^*$ in $S^d(A\ot B\ot C)^*$.
The integers $k_{\pi,\mu,\nu}$ are called {\it Kronecker coefficients} and can be computed combinatorially.
The programs \textsc{Schur} or \textsc{Sage} or several other ones will compute them for you in small cases.
We used a program written by Harm Derksen, which is based on characters of the symmetric group.

There is a simple formula for $\tdim S_{\pi}A^*$, namely\todo{}{replaced $k$ with $\aaa$. Correct?}
$$
\tdim S_{(p_1\hd p_{\aaa})}A^*=\Pi_{1\leq i<j\leq \aaa}\frac{ p_i-p_j+j-i}{j-i}
$$
see, e.g.,  \cite[Thm 6.3]{FH}.
We will be interested in cases where the dimension is small.

Let $\FS_d$ denote the group of permutations of $d$ elements.
If $\aaa=\bbb=\ccc$, then $\s_{r,\aaa,\aaa,\aaa}$ is also invariant under the
$\FS_3$-action permuting the vector spaces. Thus anytime $S_{\pi_1}A^*\ot S_{\pi_2}B^*\ot S_{\pi_3}C^*$ is in the ideal of $\s_r$,
the module  $S_{\pi_{\s(1)}}A^*\ot S_{\pi_{\s(2)}}B^*\ot S_{\pi_{\s(3)}}C^*$ will be as well,  for any $\s\in \FS_3$.

\subsection{First algorithm: to obtain a sample collection of polynomials}
\label{subsec:firstalgo}
What follows is an \todo{algorithm  \blue{from \cite{MR2097214,MR2836258}} to compute a basis}
{cite Luke Oeding if this algorithm appears in his work! JM should know.} of highest weight vectors
for each isotypic component in $S^d(A\ot B\ot C)^*$.
Once one has these, for each isotypic component, one can test if there are modules in the ideal of $\s_{r;\aaa,\bbb,\ccc}$
(or any $G$-variety for that matter) by sampling random
points on $\s_{r;\aaa,\bbb,\ccc}$ as described in the second algorithm.
For $\s\in \FS_d$, we write $\s(v_1\otc v_d):=v_{\s(1)}\otc v_{\s(d)}$.
Once and for all fix bases $a^1\hd a^{\aaa}$ of $A^*$, and similarly for $B,C$.
Let  $\pi=(p_1\hd p_{\ell},0\hd 0)$ be a partition as above. Write $\pi=(p_1\hd p_{\ell})$ and  $\ell(\pi)=\ell$.  Define
$F_{A,\pi}  \in A^{*\ot d}$
by
$$F_{A,\pi} :=(a^1)^{\ot (p_1-p_2)}\ot (a^1\ww a^2)^{\ot (p_2-p_3)}\otc (a^1\ww \cdots\ww  a^{f})^{\ot (p_{f}-p_{f-1})}.
$$
Here
$v_1\ww\cdots \ww v_k:=\frac 1{k!}\sum_{\s\in \FS_k}\tsgn(\s) \s(v_1 \otc v_k)$.

\medskip
\hrule
\begin{algorithmic}[1]
\REQUIRE Degree $d$ and partitions $\pi,\mu,\nu$ of $d$.
\ENSURE A basis $P$ of the highest weight space vector of the isotypic component of $S_{\pi}A^*\ot S_{\mu}B^*\ot S_{\nu}C^*$ in $S^d(A\ot B\ot C)^*$.
\STATE Use your favorite method to compute $k_{\pi,\mu,\nu}$.
\STATE Set $k=0$.
\WHILE{$k<k_{\pi,\mu,\nu}$}
\REPEAT
\STATE Choose permutations $\t_1,\t_2\in \FS_d$. %(Sometimes it will be possible to take $\t_1,\t_2=Id$.)
\STATE Define
$$
F_{\pi,\mu,\nu}^{\t_1,\t_2}:=F_{A,\pi} \ot (\t_1\cdot F_{B,\mu}) \ot (\t_2\cdot F_{C,\nu})
\in A^{*\ot d}\ot B^{*\ot d}\ot C^{*\ot d},
$$
rearrange the factors so it is expressed as an element of
$(A\ot B\ot C)^{*\ot d}$,
and   symmetrize to get
\begin{align*}
P_{\pi,\mu,\nu}^{\t_1,\t_2}:&=
\sum_{\s\in \FS_d}\s\cdot F_{\pi,\mu,\nu}^{\t_1,\t_2}\\
&=
\sum_{\s\in \FS_d} (\s\cdot F_{A,\pi}) \ot ( \s\cdot\t_1\cdot F_{B,\mu}) \ot
(\s\cdot\t_2\cdot F_{C,\nu})
\in A^{*\ot d}\ot B^{*\ot d}\ot C^{*\ot d}
\end{align*}
where recall  $\s\cdot (a^1\otc a^d):=a^{\s(1)}\otc a^{\s(d)}$.
% In practice, the symmetrization step just amounts to replacing $\otimes$ with a product.  ciken: We do not do this.
\UNTIL{$P_{\pi,\mu,\nu}^{\t_1,\t_2}$ is linearly independent of $P_1\hd P_{k-1}$.}
\STATE Increase $k = k+1$.
\STATE Set $P_k = P_{\pi,\mu,\nu}^{\t_1,\t_2}$.
\ENDWHILE
\end{algorithmic}
\hrule

% \smallskip
%
% \donote{Input:} Degree $d$ and partitions $\pi,\mu,\nu$ of $d$. Use your favorite method to compute $k_{\pi,\mu,\nu}$.
%
% \begin{itemize}
% \item  Set $k=0$.
%
% \item (BW)  Begin while:
% while $k<k_{\pi,\mu,\nu}$,
%
% \item (CP) choose permutations $\t_1,\t_2\in \FS_d$. (Sometimes it will be possible
% to take $\t_1,\t_2=Id$.)
%
% \item Define
% $$
% F_{\pi,\mu,\nu}^{\t_1,\t_2}:=F_{A,\pi} \ot \t_1\cdot F_{B,\mu} \ot \t_2\cdot F_{C,\nu}
% \in A^{*\ot d}\ot B^{*\ot d}\ot C^{*\ot d},
% $$
% rearrange the factors so it is expressed as an element of
% $(A\ot B\ot C)^{*\ot d}$,
% and   symmetrize to get
% \begin{align*}
% P_{\pi,\mu,\nu}^{\t_1,\t_1}:&=
% \sum_{\s\in \FS_d}\s\cdot F_{\pi,\mu,\nu}^{\t_1,\t_2}\\
% &=
% \sum_{\s\in \FS_d} \s\cdot F_{A,\pi} \ot \s\cdot\t_1\cdot F_{B,\mu} \ot
% \s\cdot\t_2\cdot F_{C,\nu}
% \in A^{*\ot d}\ot B^{*\ot d}\ot C^{*\ot d}
% \end{align*}
% where recall  $\s\cdot (a^1\otc a^d):=a^{\s(1)}\otc a^{\s(d)}$.
% In practice, the symmetrization  step just amounts to replacing $\otimes$ with a product.
%
% \item
% If $P_{\pi,\mu,\nu}^{\t_1,\t_2}=0$, re-choose $\t_1,\t_2$.
%
% \item  If  $P_{\pi,\mu,\nu}^{\t_1,\t_2}$ is not  independent of $P_1\hd P_{k-1}$,
% return to (CP),
% \item  If $k=k_{\pi,\mu,\nu}$ then done.
% \item otherwise:
%  call $P_{\pi,\mu,\nu}^{\t_1,\t_1}=:P_k$, set $k=k+1$
% return to (BW)
% \end{itemize}
%
% \donote{Output:} a basis of the highest weight space of the isotypic component of $S_{\pi}A^*\ot S_{\mu}B^*\ot S_{\nu}C^*$ in $S^d(A\ot B\ot C)^*$.

\subsubsection{Examples}\label{subsubsec:exa}
\paragraph{First example, $d=2$, $(\pi,\mu,\nu)=((2),(1,1),(1,1))$}\mbox{~}

Here $k_{(2),(1,1),(1,1)}=1$, so we are looking for a single polynomial.
We have  $F_{A,(2)} =(a^1)^2$, $F_{B,(1,1)} =b^1\ww b^2$ and
$F_{C,(1,1)} =c^1\ww c^2$.
Try $\t_1=\t_2=Id$, then
\begin{align*}
F^{Id,Id}_{(2),(1,1),(1,1)}&=(a^1\ot a^1)\ot (b^1\ot b^2- b^2\ot b^1)\ot (c^1\ot c^2- c^2\ot c^1)\\
&=(a^1\ot b^1\ot c^1)\ot(a^1\ot b^2\ot c^2) -(a^1\ot b^1\ot c^2)\ot (a^1\ot b^2\ot c^1)\\
&\ -
 (a^1\ot b^2\ot c^1)\ot(a^1\ot b^1\ot c^2) +(a^1\ot b^2\ot c^2)\ot (a^1\ot b^1\ot c^1)
 \end{align*}
 Thus
 $$
 P^{Id,Id}_{(2),(1,1),(1,1)}(x^{ijk}a_i\ot b_j\ot c_k)=
 2 x^{111}x^{122}-2 x^{112}x^{121}
 $$
Here, and throughout, repeated indices are to be summed over.
 Note that if $T=x^{ijk}a_i\ot b_j\ot c_k$ has rank one, then $P^{Id,Id}_{(2),(1,1),(1,1)}(T)=0$, but $P$ will
 evaluate to be nonzero on a general rank two tensor.

\paragraph{Second example, $d=3$, $(\pi,\mu,\nu)=((2,1),(2,1),(2,1))$}\mbox{~}

Here $k_{(2,1),(2,1),(2,1)}=1$, so again  we are looking for a single polynomial.
We have  $F_{A,(2,1)} = a^1\ot ( a^1\ww a^2)$, and similarly for $B,C$.
Try $\t_1=\t_2=Id$, then
 \begin{align*}
 &F^{Id,Id}_{(21),(21),(21)}
 =(a^1\ot a^1\ot \a^2-a^1\ot a^2\ot a^1 )\ot (b^1\ot b^1\ot b^2-b^1\ot b^2\ot b^1 ) \ot (c^1\ot c^1\ot c^2-c^1\ot c^2\ot c^1 )\\
 &=(a^1\ot b^1\ot c^1)\ot(a^1\ot b^1\ot c^1)\ot(a^2\ot b^2\ot c^2) -(a^1\ot b^1\ot c^1)\ot (a^1\ot b^1\ot c^2) \ot (a^2\ot b^2\ot c^1) \\
 &-(a^1\ot b^1\ot c^1)\ot (a^1\ot b^2\ot c^1) \ot (a^2\ot b^1\ot c^2) +(a^1\ot b^1\ot c^1)\ot(a^1\ot b^2\ot c^2)\ot(a^2\ot b^1\ot c^1)\\
 &-(a^1\ot b^1\ot c^1)\ot (a^2\ot b^1\ot c^1) \ot (a^1\ot b^2\ot c^2) +(a^1\ot b^1\ot c^1)\ot (a^2\ot b^1\ot c^2) \ot (a^1\ot b^2\ot c^1)\\
 &+(a^1\ot b^1\ot c^1)\ot(a^2\ot b^2\ot c^1)\ot(a^1\ot b^1\ot c^2) -(a^1\ot b^1\ot c^1)\ot (a^2\ot b^2\ot c^2)\ot (a^1\ot b^1\ot c^1)
  \end{align*}
 Thus
 $
 P^{Id,Id}_{(21),(21),(21)}\equiv
 0
 $
 so we need to try different $\t_1,\t_2$. Take $\t_1=Id$ and $\t_2=(12)$. Then
 \begin{align*}
F^{Id,(1,2)}_{(21),(21),(21)} &=(a^1\ot a^1\ot a^2 -a^1\ot a^2\ot a^1 )\ot (b^1\ot b^1\ot b^2-b^1\ot b^2\ot b^1)
\ot (c^1\ot c^1\ot c^2-c^2\ot c^1\ot c^1 )
\\
&=(a^1\ot b^1\ot c^1)\ot(a^1\ot b^1\ot c^1)\ot(a^2\ot b^2\ot c^2) -(a^1\ot b^1\ot c^2)\ot (a^1\ot b^1\ot c^1)
\ot (a^2\ot b^2\ot c^1)
\\
& -(a^1\ot b^1\ot c^1)\ot (a^1\ot b^2\ot c^1)\ot (a^2\ot b^1\ot c^2)
+(a^1\ot b^1\ot c^2)\ot (a^1\ot b^2\ot c^1)\ot (a^2\ot b^1\ot c^1)
\\
& -(a^1\ot b^1\ot c^1)\ot (a^2\ot b^1\ot c^1)\ot (a^1\ot b^2\ot c^2)
 +(a^1\ot b^1\ot c^2)\ot (a^2\ot b^1\ot c^1)\ot (a^1\ot b^2\ot c^1)
 \\
 &+
 (a^1\ot b^1\ot c^1)\ot(a^2\ot b^2\ot c^1)\ot(a^1\ot b^1\ot c^2) -(a^1\ot b^1\ot c^2)\ot (a^2\ot b^2\ot c^1)
 \ot (a^1\ot b^1\ot c^1).
 \end{align*}
 Thus $\displaystyle P^{Id,(1,2)}_{(21),(21),(21)}\left(\sum_{i,j,k=1}^2 x^{ijk}a_i\ot b_j\ot c_k\right)=$
\[
%   x^{111}x^{111} x^{222}   +2x^{112}x^{121} x^{211}
%   - x^{111}x^{121} x^{212}
%     -x^{111}x^{211} x^{122}
%       - x^{111}x^{112}x^{221}.
\red{  x^{111}x^{111} x^{222}   +2x^{112}x^{121} x^{211}
  - (
  x^{111}x^{121} x^{212}
    +x^{111}x^{211} x^{122}
       +x^{111}x^{112}x^{221}).}
\]\todo{}{I think the formula above is correct although the reviewer is suspicious. I added brackets to highlight the structure.}

Note that if $T$ has rank one, then $P^{Id,(1,2)}_{(21),(21),(21)}(T)=0$, but $P$ will
 evaluate to be nonzero on a general rank two tensor.

% Exercise: take $(\pi,\mu,\nu)=((21),(21),(111))$, find the corresponding polynomial, and show it is in $I_3(\s_2)$.

\subsubsection{Permutation pairs to avoid}
We want to avoid the case that occurs in the first try of \magenta{the} \green{second} example of~Section~\ref{subsubsec:exa}, i.e., that $P_{\pi,\mu,\nu}^{\tau_1,\tau_2}=0$.
Although a complete classification of the cases when this happens is unknown, an easy necessary condition for $P_{\pi,\mu,\nu}^{\tau_1,\tau_2}\neq 0$ is the following \cite[Lemma 7.2.7]{ike:12}:
When we write $1,2,\ldots,d$ in a tableau columnwise starting with the longest column and
we write $\tau_1(1),\tau_1(2),\ldots,\tau_1(d)$ in a second tableau columnwise,
and we do the same for $\tau_2$ in a third tableau,
then it is necessary that there exists no pair of numbers that lies in the same column in all three tableaux.
If this occurs, we call this situation a \emph{zero pattern}.
We can choose random permutations that avoid the zero pattern by just choosing random permutations
and repicking if it contains a zero pattern.

\red{
\subsubsection{Implementation}\label{subsubsec:implI}
\blue{T}he algorithm is not complicated to implement,
but the following details are paramount for its running time.

What is crucial in our implementation is that we avoid writing down the $P_i$ as polynomials.
A polynomial $P_i$ i\blue{s} stored only as its permutation pair $(\tau_1,\tau_2) \in \FS_d \times \FS_d$.
To prove linear independence among polynomials $P_i$, the $P_i \in S^d(A\otimes B \otimes C)^*$ are
contracted with random tensors $t_j = w_j^{\otimes d}$ with $w_j \in A \otimes B \otimes C$ having low rank,
which is the same as evaluating the function $P_i$ on $w_j$.
If the resulting matrix $(\left< P_i, t_j \right>)_{i,j}$ consisting of the contractions $\left< P_i, t_j \right>$ has full rank, then the $P_i$ are linearly independent.

If $a \in (A\otimes B \otimes C)^{\otimes d}$ is of rank 1,
then the contraction $\left< P_i, a \right>$ is a product of $\ell \times \ell$ determinants,
which can be efficiently computed.
Hence to compute a contraction $\left<P_i, t_j \right>$,
it would suffice to expand $t_j$ into rank 1 tensors and sum over the products of determinants.
But to make this method computationally feasible, we do not expand $t_j$ completely,
since $\left<P_i, t_j \right>$ would consist of a huge amount of zero summands.
We use a standard divide and conquer method to expand $t_j$ partially and prune the computation
whenever at least one determinant is seen to be zero.

To avoid numerical errors, for proving linear independence working over a finite field or ring suffices.
The same method can be used to evaluate at the matrix multiplication tensor $M_2$.
}

\subsection{Second algorithm: to test on the secant variety}
Once one has a basis of highest weight vectors for an isotypic component, one needs to
determine which linear combinations of basis vectors vanish on $\s_r$.
The \todo{following algorithm (see, e.g., \blue{from \cite{MR2097214,MR2836258}})}{Reviewer: Has
this appeared in the literature yet?} is standard linear algebra:

\medskip
\hrule
\begin{algorithmic}[1]
\REQUIRE The output of first algorithm for some $(\pi,\mu,\nu)$, i.e., a collection
$P_1\hd P_k=P_{k_{\pi,\mu,\nu}}\in S^d(A\ot B\ot C)^*$ and $r$, where we will
test for polynomials in $I(\s_{r;\aaa,\bbb,\ccc})$.
\ENSURE with probability as high as you like the component of $I(\s_{r;\aaa,\bbb,\ccc})$
in $S_{\pi}A^*\ot S_{\mu}B^*\ot S_{\nu}C^*$.  If the component is zero, then the answer
is guaranteed correct, and more generally, the algorithm can only overestimate the component
if the points on $\s_r$ are not chosen randomly enough.
\STATE Set $P=c_1 P_1+\cdots c_k P_k$, where  $c_1\hd c_k$ are variables.
\STATE Chose \lq\lq random\rq\rq vectors
$$v_j=\sum_{i=1}^{\aaa} \sum_{k=1}^{\bbb}\sum_{l=1}^{\ccc}(\a^i_{1,j} a_i)\ot (\b^k_{1,j} b_k)\ot (\g^l_{1,j} c_l)+\cdots +(\a^i_{r,j} a_i)\ot (\b^k_{r,j} b_k)\ot (\g^l_{r,j} c_l)$$ where the $\a^i_{\delta,j},\b^k_{\delta,j},\g^l_{\delta,j}$ are \lq\lq random\rq\rq\
numbers.
\STATE Evaluate $P$ at these $k$ points.
\IF{there exist a solution $ c_1\hd c_k$ such that all the evaluations are zero}
\STATE If there is a $m$-dimensional  solution space, then with reasonable probability one has $m$ copies of the module in the ideal.
\ELSE
\STATE No module in this isotypic component is in $I(\s_{r;\aaa,\bbb,\ccc})$.
\ENDIF
\end{algorithmic}
\hrule

\red{
\subsubsection{Implementation}
Again, here it is crucial to store the $P_i$ only as permutation pairs.
Evaluation at points works as described in Section~\ref{subsubsec:implI}.
Unlike linear independence, we need stronger methods to prove linear dependence.
One can parametrize $\s_6$ and use the fact that the relations between
all determinants that appear during the calculation are given by Young tableau relations, cf.~\cite[p.~110]{fult:97}.
No particular optimization was done during this step, which renders it the slowest part of our algorithm.
}

\subsection{Our run}
\red{
Let $d=19$, $(\pi,\mu,\nu)=((5554),(5554),(5554))$.
Here $k_{(5554),(5554),(5554)}=31$.
We found $31$ pairs $\t_1,\t_2$ that result in 31 linearly independent polynomials by
choosing $\t_1$ and $\t_2$ randomly, but avoiding the zero pattern.
As expected, the linear combination has support $31$ and no evident structure other than
that it ``magically'' vanishes on $\s_{6;4,4,4}$.
A somewhat nicer description (smaller support) of a polynomial vanishing on $\s_{6;4,4,4}$
is obtained in the following $d=20$ case.}

Let $d=20$, $(\pi,\mu,\nu)=((5555),(5555),(5555))$.
Here $k_{(5555),(5555),(5555)}=4$.
The following \red{random} choices of pairs $\t_1,\t_2$ give $4$ linearly independent polynomials.\\
$\tau_1 = (\tau_1(1),\tau_1(2),\ldots,\tau_1(20)) = (10, 15, 5, 9, 13, 4, 17, 14, 7, 20, 19, 11, 2, 12, 8, 3, 16, 18, 6, 1)$,\\
$\tau_2 = (10, 11, 6, 2, 8, 9, 4, 20, 15, 16, 13, 18, 14, 19, 7, 5, 17, 3, 12, 1)$

\smallskip

\noindent$\tau_1 = (19, 10, 1, 5, 7, 12, 2, 13, 16, 6, 18, 9, 11, 20, 3, 17, 14, 8, 15, 4)$,\\
$\tau_2 = (10, 5, 13, 6, 3, 16, 11, 1, 4, 18, 15, 17, 9, 2, 8, 12, 19, 7, 14, 20)$

\smallskip

\noindent$\tau_1 = (16, 20, 9, 13, 8, 1, 4, 19, 11, 17, 7, 2, 14, 3, 6, 5, 12, 15, 18, 10)$,\\
$\tau_2 = (1, 20, 11, 19, 5, 16, 17, 2, 18, 13, 7, 12, 14, 10, 8, 15, 6, 9, 3, 4)$

\smallskip

\noindent$\tau_1 = (11, 5, 2, 1, 16, 10, 20, 3, 17, 19, 12, 18, 13, 9, 14, 4, 8, 6, 15, 7)$,\\
$\tau_2 = (1, 6, 15, 13, 20, 3, 18, 11, 14, 2, 9, 5, 4, 17, 12, 8, 19, 16, 7, 10)$

\smallskip

\noindent This gives rise to 4 polynomials $f_1,\ldots,f_4$.
\red{Restricting these functions to $\sigma_6$, the second algorithm yields} the following linear combination, which vanishes on~$\sigma_6$:
$-266054\,f_1 + 421593\,f_2 + 755438\,f_3 + 374660\,f_4$.
\red{The coefficients look \blue{r}andom, as is expected, since the permutation pairs were chosen at random.
The computation took several hours on 16 processors, the symbolic proof of vanishing at $\sigma_6$ being by far the slowest part.}

\strike{\begin{remark} One can parametrize $\s_r$ and in
some cases  use the explicit parametrization to get an exact answer, which is
what we did   for $(\pi,\mu,\nu)=((5555),(5555),$ $(5555))$.
\end{remark}}

% \donote{Input:} Output of first algorithm for some $(\pi,\mu,\nu)$, a collection
% $P_1\hd P_k=P_{k_{\pi,\mu,\nu}}\in S^d(A\ot B\ot C)^*$ and $r$, where we will
% test for polynomials in $I(\s_{r;\aaa,\bbb,\ccc})$.
%
% \begin{itemize}
%
% \item
% Set $P=c_1 P_1+\cdots c_k P_k$, where  $c_1\hd c_k$ are variables.
% Chose \lq\lq random\rq\rq vectors
% ....
%
% Evaluate $P$ at these $k$ points.
%
% \item Does there exist a solution $ c_1\hd c_k$ such that
% all the evaluations are zero?
%
%  \item If no solution: done - no module in this isotypic component is in $I(\s_{r;\aaa,\bbb,\ccc})$.
%
% \item If there is a $m$-dimensional  solution space, then with reasonable probability one has
% $m$ copies of the module in the ideal.
%
% \end{itemize}
%
%
% Output: with probability as high as you like the component of $I(\s_{r;\aaa,\bbb,\ccc})$
% in $S_{\pi}A^*\ot S_{\mu}B^*\ot S_{\nu}C^*$.  If the component is zero, then the answer
% is guaranteed correct, and more generally, the algorithm can only overestimate the component
% if the points on $\s_r$ are not chosen randomly enough. One can parametrize $\s_r$ and in
% small dimensions use the explicit parametrization to get an exact answer, which is
% what we did in ***.\randbem{I did the symbolic calculation for $(\pi,\mu,\nu)=((5555),(5555),$ $(5555))$, if you mean this.}

\section{Review of the original proof that $M_2\not\in\s_{6;4,4,4}$}\label{Sec:ReviewProof}

The essence of the proof that the \green{border} rank of $M_2$ is not six in \cite{Lmatrix} is as follows:
there is a now standard argument due to Baur for proving lower bounds for rank by splitting
a putative computation into two parts using the algebra structure on the space of matrices.
  The argument in \cite{Lmatrix} was to
apply the same type of argument  to each component of the variety consisting of subvarieties where the rank is greater than the border rank.
The article \cite{Lmatrix} contained a gap in the proof that was filled in \cite{Lmatrixarxiv} but not published in JAMS because
the editor was concerned the erratum was almost as long as the original article and the author did not see a way to shorten it.
The gap in \cite{Lmatrix} was caused by overlooking the possibility of certain types of components, where the limiting $6$-planes are not
formed by points coming together but by some other unusual configuration of points. All such components of $\s_{6;4,4,4}$ are not
known explicitly, but the correction only used qualitative aspects of how the limiting $6$-plane arose. There were $3$ basic cases,
where any subset of $5$ of the limit points are linearly independent, where there is a subset of $5$ that are not, but any subset of
four are, and where there is a subset of $4$ that are not, but any subset of $3$ are. In each of these cases, one is forced to
have a limit taking place among rank one tensors in a much smaller space, which was what made the analysis tractable.
The computations performed above provide an explicit polynomial
vanishing on $\s_{6;4,4,4}$ which does not vanish at $M_2$, providing  a significantly shorter proof
of this~fact. %jml 4-26

\bibliographystyle{amsplain}

\bibliography{HIL}

\def\cdprime{$''$} \def\cprime{$'$} \def\cprime{$'$} \def\cprime{$'$}
  \def\Dbar{\leavevmode\lower.6ex\hbox to 0pt{\hskip-.23ex \accent"16\hss}D}
  \def\cprime{$'$} \def\cprime{$'$} \def\cdprime{$''$} \def\cprime{$'$}
  \def\cprime{$'$} \def\Dbar{\leavevmode\lower.6ex\hbox to 0pt{\hskip-.23ex
  \accent"16\hss}D} \def\cprime{$'$} \def\cprime{$'$} \def\cprime{$'$}
  \def\cprime{$'$} \def\Dbar{\leavevmode\lower.6ex\hbox to 0pt{\hskip-.23ex
  \accent"16\hss}D} \def\cprime{$'$} \def\cprime{$'$}
\providecommand{\bysame}{\leavevmode\hbox to3em{\hrulefill}\thinspace}
\providecommand{\MR}{\relax\ifhmode\unskip\space\fi MR }
\providecommand{\MRhref}[2]{%
  \href{http://www.ams.org/mathscinet-getitem?mr=#1}{#2}
}
\providecommand{\href}[2]{#2}
\begin{thebibliography}{10}

\bibitem{batesoeding}
D.~Bates and L.~Oeding, \emph{Toward a salmon conjecture}, preprint,
  arXiv:1009.6181.

\bibitem{Bertini}
Daniel~J Bates, Jonathan~D Hauenstein, Andrew~J Sommese, and Charles~W Wampler,
  \emph{Bertini: Software for numerical algebraic geometry}, 2006.

\bibitem{MR2836258}
Daniel~J. Bates and Luke Oeding, \emph{Toward a salmon conjecture}, Exp. Math.
  \textbf{20} (2011), no.~3, 358--370. \MR{2836258 (2012i:14056)}

\bibitem{BHORS}
G.~Blekherman, J.~Hauenstein, J.C. Ottem, K.~Ranestad, and B.~Sturmfels,
  \emph{Algebraic boundaries of {H}ilbert's {SOS} cones}, Compositio
  Mathematica \textbf{148} (2012), 1717--1735.

\bibitem{BCS:97}
Peter B{\"u}rgisser, Michael Clausen, and M.~Amin Shokrollahi, \emph{Algebraic
  complexity theory}, Grundlehren der Mathematischen Wissenschaften, vol. 315,
  Springer-Verlag, Berlin, 1997, With the collaboration of Thomas Lickteig.
  \MR{1440179 (99c:68002)}

\bibitem{BIrank}
Peter B{\"u}rgisser and Christian Ikenmeyer, \emph{Geometric {C}omplexity
  {T}heory and {T}ensor {R}ank}, Proceedings 43rd Annual ACM Symposium on
  Theory of Computing 2011 (2011), 509--518.

\bibitem{BItensor}
\bysame, \emph{Explicit {L}ower {B}ounds via {G}eometric {C}omplexity
  {T}heory}, Proceedings 45th Annual ACM Symposium on Theory of Computing 2013
  (2013), 141--150.

\bibitem{fult:97}
William Fulton, \emph{Young tableaux}, London Mathematical Society Student
  Texts, vol.~35, Cambridge University Press, Cambridge, 1997, With
  applications to representation theory and geometry. \MR{1464693 (99f:05119)}

\bibitem{FH}
William Fulton and Joe Harris, \emph{Representation theory}, Graduate Texts in
  Mathematics, vol. 129, Springer-Verlag, New York, 1991, A first course,
  Readings in Mathematics. \MR{1153249 (93a:20069)}

\bibitem{GHPS}
Z.A. Griffin, J.D. Hauenstein, C.~Peterson, and A.J. Sommese, \emph{Numerical
  computation of the {H}ilbert function of a zero-scheme}, Available at {\tt
  www.math.ncsu.edu/$\sim$jdhauens/preprints}.

\bibitem{HHM12}
J.D. Hauenstein, Y.-H. He, and D.~Mehta, \emph{Numerical analyses on moduli
  space of vacua}, {\tt arXiv:1210.6038}, 2012.

\bibitem{HS10}
J.D. Hauenstein and A.J. Sommese, \emph{Witness sets of projections}, Appl.
  Math. Comput. \textbf{217} (2010), no.~7, 3349--3354.

\bibitem{HS13}
\bysame, \emph{Membership tests for images of algebraic sets by linear
  projections}, 2013, pp.~6809--6818.

\bibitem{ike:12}
Christian Ikenmeyer, \emph{Geometric {C}omplexity {T}heory, {T}ensor {R}ank,
  and {L}ittlewood-{R}ichardson {C}oefficients}, Ph.D. thesis, Institute of
  Mathematics, University of Paderborn, 2012, Online available at
  \url{http://math-www.uni-paderborn.de/agpb/work/ikenmeyer_thesis.pdf}.

\bibitem{Lmatrixarxiv}
J.~M. Landsberg, \emph{The border rank of the multiplication of {$2\times2$}
  matrices is seven}, arXiv:math/0407224.

\bibitem{Lhighbranktensor}
\bysame, \emph{Explicit tensors of border rank at least 2n-1}, preprint
  arXiv:1209.1664.

\bibitem{Lmatrix}
\bysame, \emph{The border rank of the multiplication of {$2\times2$} matrices
  is seven}, J. Amer. Math. Soc. \textbf{19} (2006), no.~2, 447--459
  (electronic). \MR{2188132 (2006j:68034)}

\bibitem{MR2865915}
\bysame, \emph{Tensors: geometry and applications}, Graduate Studies in
  Mathematics, vol. 128, American Mathematical Society, Providence, RI, 2012.
  \MR{2865915}

\bibitem{MR2097214}
J.~M. Landsberg and L.~Manivel, \emph{On the ideals of secant varieties of
  {S}egre varieties}, Found. Comput. Math. \textbf{4} (2004), no.~4, 397--422.
  \MR{2097214 (2005m:14101)}

\bibitem{LMsec}
J.~M. Landsberg and Laurent Manivel, \emph{On the ideals of secant varieties of
  {S}egre varieties}, Found. Comput. Math. \textbf{4} (2004), no.~4, 397--422.
  \MR{2097214 (2005m:14101)}

\bibitem{LOsecbnd}
J.M. Landsberg and Giorgio Ottaviani, \emph{New lower bounds for the border
  rank of matrix multiplication}, preprint, arXiv:1112.6007.

\bibitem{Migliore}
J.C. Migliore, \emph{Introduction to liaison theory and deficiency modules},
  Progress in Mathematics, vol. 165, Birkh\"auser Boston Inc., Boston, MA,
  1998.

\bibitem{MR623057}
A.~Sch{\"o}nhage, \emph{Partial and total matrix multiplication}, SIAM J.
  Comput. \textbf{10} (1981), no.~3, 434--455. \MR{623057 (82h:68070)}

\bibitem{Monodromy}
A.J. Sommese, J.~Verschelde, and C.W. Wampler, \emph{Using monodromy to
  decompose solution sets of polynomial systems into irreducible components},
  Applications of algebraic geometry to coding theory, physics and computation
  ({E}ilat, 2001), NATO Sci. Ser. II Math. Phys. Chem., vol.~36, Kluwer Acad.
  Publ., Dordrecht, 2001, pp.~297--315.

\bibitem{Trace}
\bysame, \emph{Symmetric functions applied to decomposing solution sets of
  polynomial systems}, SIAM J. Numer. Anal. \textbf{40} (2002), no.~6,
  2026--2046.

\bibitem{SW05}
A.J. Sommese and C.W. Wampler, II, \emph{The numerical solution of systems of
  polynomials}, World Scientific Publishing Co. Pte. Ltd., Hackensack, NJ,
  2005, Arising in engineering and science.

\bibitem{str:69}
Volker Strassen, \emph{Gaussian elimination is not optimal}, Numer. Math.
  \textbf{13} (1969), 354--356.

\end{thebibliography}

\end{document}